\documentclass[
aps,%
12pt,%
final,%
notitlepage,%
oneside,%
onecolumn,%
nobibnotes,%
nofootinbib,%
superscriptaddress,%
noshowpacs,%
centertags]%
{revtex4}
\usepackage{aas_macros}
\begin{document}
\selectlanguage{russian}
                                \begin{center}
                                
                                 {\bf  }
\end{center}

\title{CHANGING LOOKS OF THE NUCLEUS OF SEYFERT GALAXY NGC~3516 DURING 2016-2020}

\author{\firstname{V.~L.}~\surname{Oknyansky}}
\email{oknyan@mail.ru}
\affiliation{%
Sternberg  Astronomical Institute, M. V. Lomonosov Moscow State University, 119234, Moscow, Universitetsky pr-t, 13, Russian Federation
}%
\author{\firstname{Kh.~M.}~\surname{Mikailov}}
\affiliation{%
Shamakhy Astrophysical Observatory, National Academy of Sciences, Baku, Azerbaijan
}%

\author{\firstname{N.~A.}~\surname{Huseynov}}
\affiliation{%
Shamakhy Astrophysical Observatory, National Academy of Sciences, Baku, Azerbaijan
}%


\begin{abstract}

The results of spectral observations of NGC~3516  with the 2-m telescope of the Shamakhy Astrophysical Observatory during 2016-2020 are presented. In the first half of 2016, the intensive broad component H${\beta}$ was found, which indicates a  spectral type change compared to 2014, when the broad component was almost invisible. In the second half of 2016, the  broad component H${\beta}$ again was weakened and was practically not observed, remaining as weak until the end of 2019. At the end of 2019, the broad component $H{\beta}$ strengthened again, and in May 2020 reached a typical level for the high state of the object. During 2016-2020 we observed several changing looks of  NGC~3516.
\end{abstract}

\maketitle

\section{Introduction}

NGC~3516 was discovered in 1785 (William Herschel) and is included in the classic list
of K.Seyfert \cite{se43}.  It is the first discovered Seyfert galaxy with 
spectral variability (1968) and, at the same time, the first discovered active
galactic nucleus (AGN), which changes its spectral type \cite{as68, sh19}.  However
in 1968, the erroneous assumption was made that variables were forbidden
lines, but the continuum and Balmer lines did not change. The discovery of the 
variability of the H${\alpha}$  in NGC~3516 and, at the same time, the first measurement of the
time delay in this line variations relative to the continuum (for AGN as a whole) was 
done in 1973 \cite{Ch1973}.  AGN classification by type is based on the properties of emission 
lines. The first type (Sy1) includes objects which have broad
lines (with half-width at half-intensity (HWHM) more than 1000 km/s) and
much narrower forbidden lines with HWHM less than 1000 km/s.  The second type
 (Sy2) includes objects in which permitted and forbidden lines have
approximately the same HWHM, less than 1000 km/s  (see details and links, for 
example, in \cite{di81}).  From the point of view of the unification model (UM), these spectral 
differences \cite{an93} explains by the presence of a dust torus and its different orientation 
with respect to the observer.
  The existence of AGN changing their type in a relatively short time (months
or less) is a serious problem for the UM.  Several dozen changing their type AGNs are known by present time and they got the name Changing Look AGN (CL AGN).  The nucleus of NGC~3516 has significant amplitude of variability and 
intensively observed  during last 50 years (see the survey in \cite{sh19, Ilic2020}.  After a maximum in 2006, the object's luminosity decreased, and in 2012-2015  the object was in a deep 
minimum.  In 2014, was recorded CL event in NGC~3516, when the broad H${\beta}$  was 
practically invisible \cite{sh19}.  In the end of 2015 - the first half of 2016, according to 
photometric observations, object had brightening \cite{sh19, Ilic2020}.  Unfortunately, there are no published spectra of the object in 2016 In the second half of 2016, the object was weakened 
and was in a  low state  until the end of 2018 (see, e.g., \cite{sh19, Ilic2020}), and 
then the awakening activity of the object in 2019 and a significant outburst in the 
continuum (in X-rays, UV and optical range) at the beginning of 2020, accompanied by a 
new CL, was discovered  \cite{ok20}.  We included nucleus of the NGC~3516  in the list of 
objects for our CL AGN monitoring project \cite{Oknyansky2016}.  Our spectral observations fall on the interval in 2016-2020, when significant changes occurred in NGC~3516, and other 
spectral observations (according to publications) were absent or few in number.  
Description of these observations and results are given below in the following 
sections.

\section{Spectral observations}
Optical spectral  observations  of the galaxy  NGC~3516  were carried out during 2016-2019 at 2-m telescope of Shamakhy Astropyisical Observatory within 11 nigths.  To obtain the spectra, three different spectrographs were used.: (1) - 2x2 prism 
spectrograph with FLI CCD cameras (4096x4096, 1 pixel = 9${\mu}$)\cite{m2014};  (2) - UAGS + with Uranus lens (f = 250 mm, f/2.5) + FLI CCD camera (3056x2048,
1 pixel = 9${\mu}$));  (3) - UAGS + Canon EF lens (f = 200mm, f/2) + Andor CCD camera
 (ikonL-936-BEX2-DD 2048x2048, 1 pixel = 13.5${\mu})$) ).  The wavelength range was from
3800\AA~ to 8000\AA~, and the spectral resolution varied accordingly between (3.8--5.3)\AA~ and (8.3--10.6)\AA. Signal-to-noise ratio (S/N) in the continuum near the H${\beta}$ 
line was 25--130.  Spectrophotometric standard stars were observed every night.  The 
first spectrograph was operated for a long time in the classical version with the 
photographical registration of spectra (see, e.g.,\cite{Rustamov2001}), and then
was upgraded to record spectra using a CCD device \cite{m2014} .  This
spectrograph was mainly used to obtain low-dispersion spectra of the
faint non-stationary stars.  The estimates made for the  accuracy of 
equivalent widths of absorption lines (less than 10\%), of course, are of little use for 
assessing the accuracy  of broad emission lines measurement in AGN spectra, but can 
serve as a lower estimate of the magnitude of the errors.  The investigations  of AGN  carried out  with this spectrograph  were published  only  in \cite{Oknyansky2017a, Oknyansky2017b}.  Spectrographs (2) and (3) began to 
be used only in the last few years, and therefore there are few publications on them 
(see, for example, \cite{Ts2019, Ismailov2020}).

Carrying out  of the spectra  was realized with the help of  the new version  of the   DECH software package. 
All spectra were extracted using an IRAF mask.  The process included
dark current subtraction, flat field corrections, cosmic ray removal, 2D
linearization of wavelengths, subtraction of the sky spectrum, subtraction of the 
spectrum of the galaxy (stars
component), etc.  Table 1 provides a summary of the equipment
 and  of observations.  Spectral resolution and dispersion are indicated for the region
near H${\beta}$.  We measured the intensity of the H${\beta}$ line with respect to the [OIII]$ \lambda$5007 line,
which was considered as a constant during our observations.  Fig. 1 shows 
 examples of the calibrated nucleus spectra  in relative intensities,
received in 2016-2020.  Fig. 2 shows an example of a calibrated spectrum in absolute flux received on May 22, 2020.
Fig. 3 shows examples of H${\beta}$ profiles for 3  different dates.  

As it can be seen from Table 1., our spectral data are inhomogeneous in used equipment  
and quality.  Data can be conventionally divided into 2 groups: data obtained in 
2016-2018, where the quality was lower, and, the data obtained from December 2019 to 
May 2020, when data were obtained with the new spectrograph  and were  more 
homogeneous. 
 It is possible to judge on the accuracy of  measurement of H${\beta}$ intensity by its  variations  in the closest dates.      The last 3 spectra 
were obtained within 8 days and the range of variations in the intensity of H${\beta}$ was 
about 10\%, however these variations might be partly due to the real variability of the 
line.  We do, however, took the conservative estimate of 10\% as the mean square error 
of measurement lines intensities in the last four spectra.  In the first spectra 
obtained in 2016-2018, measurement errors are much bigger, which is associated with the 
features of the spectrographs, different spectral slit widths and quality of weather 
conditions.  Influence of the width of the spectral slit for measuring the intensity of H${\beta}$ 
can be estimated from data received on closest dates.  According to our estimates, the 
change in the slit width from 1" to 2" can give an error that does not exceed the 
measurement error, which  according to our estimation was about 20\% (for data obtained 
during 2016-2018).  Spectra obtained in 2 neighboring  nights in October 2017 were 
combined into one  to reduce noise.  Note that in the case of a weak H${\beta}$ line, when the 
broad component is almost invisible, errors can
be much bigger.  In \cite{sh19} estimates  of the effect of the slit width on the 
measured the intensity Hb were investigated and the variations did not exceed 30\%
 when the slit was  changed by a factor of 2, however the data were obtained with different telescopes.   The size of the slit was chosen by us  depending of the image quality, what partially  reduces the role of changing conditions and the size of the slit for measuring the 
intensity of H${\beta}$.  It should be noted that errors in the measurement of equivalent  
widths of lines can be significantly bigger than in measuring of the relative intensity of 
the lines, due to the significant correlated errors associated with the procedure of 
the continuum fitting.  The variations of H${\beta}$ intensity  during 2016-2020 are shown in 
the top panel of Fig.4.

\section{Results}

As it is clearly seen from Fig. 1, the intensity of the broad component of H${\beta}$
experienced significant changes: in the spring of 2016, this line was typical for the 
Sy1, then it was weakened by August 2016. In 2017-2018  the broad 
component of H${\beta}$ was very weak, almost invisible, which is typical for Sy1.9, and at 
the end of 2019 the H${\beta}$ line began to grow again and in May 
2020 the spectrum became typical for Sy1. These conclusions 
confirm the results published earlier   \cite{Ilic2020, Oknyansky2020}. However, in 2016 no other
results of spectral observations of NGC~3516 were published. Our result on the CL event at the spring of 2016 (compared to 2014 \cite{sh19}) is new.  From photometry
series and $Swift$ data  \cite{Ilic2020} at the end of 2015 - beginning of 2016, the object had a flare and therefore, a significant enhancement of broad emission lines in spring of 2016
was connected to the increase in the flux of ionizing radiation from a central source.
The weakening of emission lines in 2017-2018 and 
brightening from the end of 2019 to the present, when a new bright
flare in the continuum  was observed \cite{Oknyansky2020}, can be explained in the same way. 

 Fig. 2 shows one of the latest spectra 
obtained on May 22, 2020.  It is clearly seen not  only the strong Balmer lines 
H${\alpha}$, H${\beta}$ and H${\gamma}$ but  also the HeII$\lambda$4681 line and the [FeX]$\lambda6374$ coronal line (see  also  \cite{Ilic2020}).  Variability of H${\beta}$ profile for three dates May 10, 2016, May 16, 2018, and May 22, 2020 is shown in Figure 3. The variability of H${\beta}$ intensity relatively to [OIII]$\lambda$5007 is shown in Fig.4.  For comparison this figure shows previously published photometric  data (see details in  \cite{Ilic2020}).  A bright flare in February-May 2020 was discovered in \cite{Oknyansky2020} and this is confirmed by our results on a significant strengthening of emission lines in May 2020. A feature of the H${\beta}$ profile in May 2020 is the appearance of a very bright
peak in the blue wing, which dominates the broad line component. The red-shifted component has also been observed, but much weaker in intensity.  However, the red component 
appears to have been present as well in 2016-2018, but the blue peak appeared and 
intensified only recently.  Thus during 2016-2020  the object was in an unstable state 
and was demonstrated several CL events.  High activity states and Sy1 spectrum 
were observed early 2016 and in May 2020.

\section{Conclusion}

We carried out spectral observations of the nucleus of the Seyfert galaxy
NGC~3516 during 2016-2020  The obtained results  showed significant variability
of H${\beta}$, which can be identified as CL events in 2016 and in
2020.  The interval of about 4 years between the high states of the NGC~3516 
nucleus is practically coincides with the found period in the variability of the object \cite{Kovacevic2020} . 
 We have compared our spectral data with published photometric data and found
the agreement between the spectral variability  and the changes in the continuum.
Thus, CL events can be interpreted as a consequence of changes in the luminosity of the 
central source.  There are several physical processes which could cause such
dramatic changes of the luminosity: variable absorption along the line of 
sight, various types of instabilities in accretion
disk, tidal destruction of stars by a super-massive black hole, close to
black hole moving of stars with partial loss  of the envelope, etc. Recurrent  
CL events in  AGN  can find a natural explanation, as in models with 
instability in the accretion disk as well as in the model with tidal stripping of stars 
which  have bound elongated orbits around the central black hole.  Required references 
and more detailed discussion of possible physical mechanisms of CL events in AGN can be found, for example, in \cite{Ilic2020, Oknyansky2017a, Oknyansky2019, Oknyansky2020,  MacLeod2019}.


\begin{center}

                      ACKNOWLEGMENTS

\end{center}
The authors express their  gratitude to the ShAO staff for their help in organizing 
and conducting spectral observations, as well as  to A.M.  Cherepashchuk and D. Ilic 
for their useful discussions.


\appendix

%
%
\bibliography{NGC3516}
%
\newpage

\begin{figure}[t!]
\includegraphics{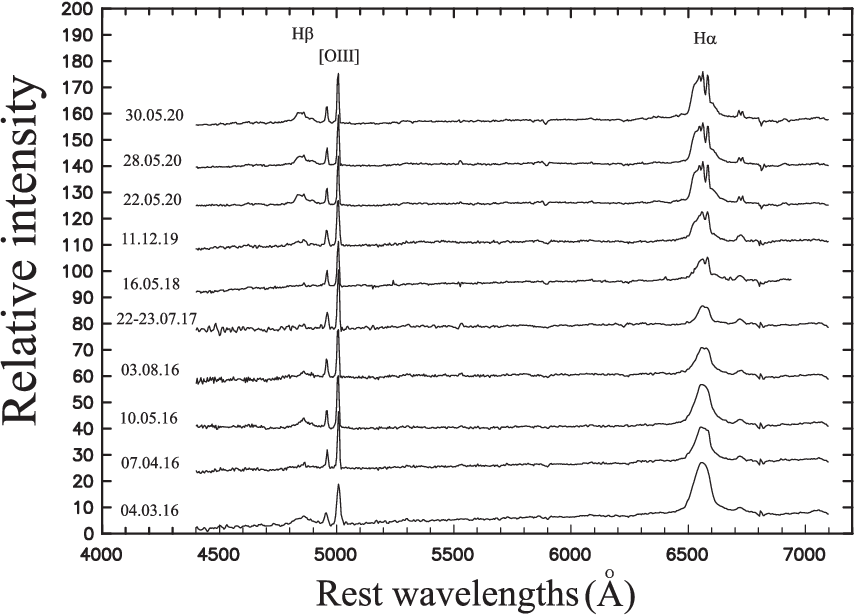}
\captionstyle{normal} \caption{Calibrated spectra of NGC~3516 in relative intensities (shifted arbitrarily along the ordinate for optimal clarity) for 2016-2020.
}
\end{figure}

\begin{figure}[t!]
\includegraphics{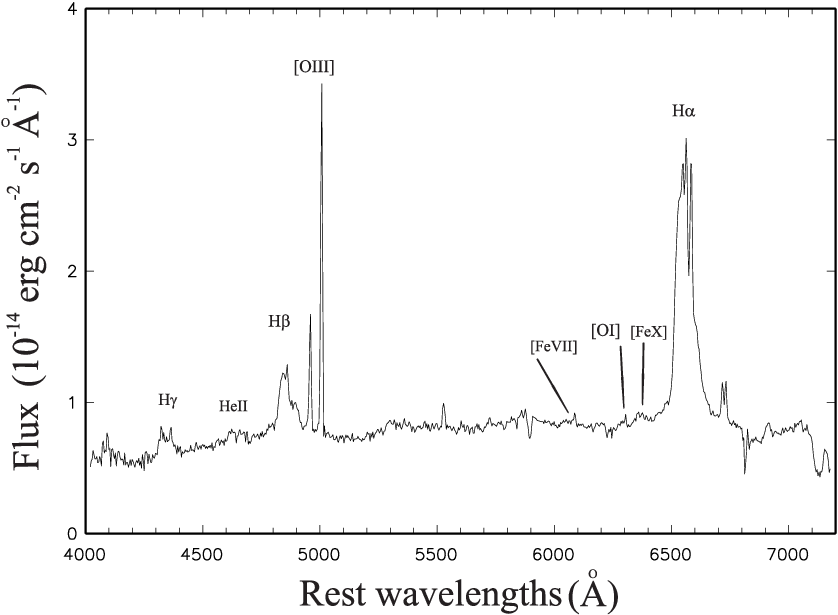}
\captionstyle{normal} \caption{The isolated nuclear non-stellar spectrum
of NGC~3516 obtained on May 22, 2020.
}
\end{figure}

\begin{figure}[t!]
\includegraphics{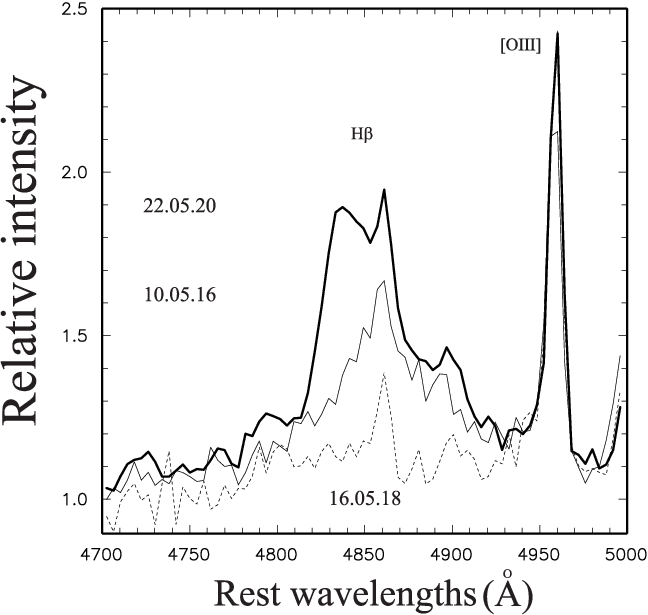}
\captionstyle{normal} \caption{Spectra of NGC~3516 in the H${\beta}$ region, normalized to the continuum at three dates:  10 May 2016, 16 May 2018  and 22 May 2020.}
\end{figure}


\begin{figure}[t!]
\includegraphics{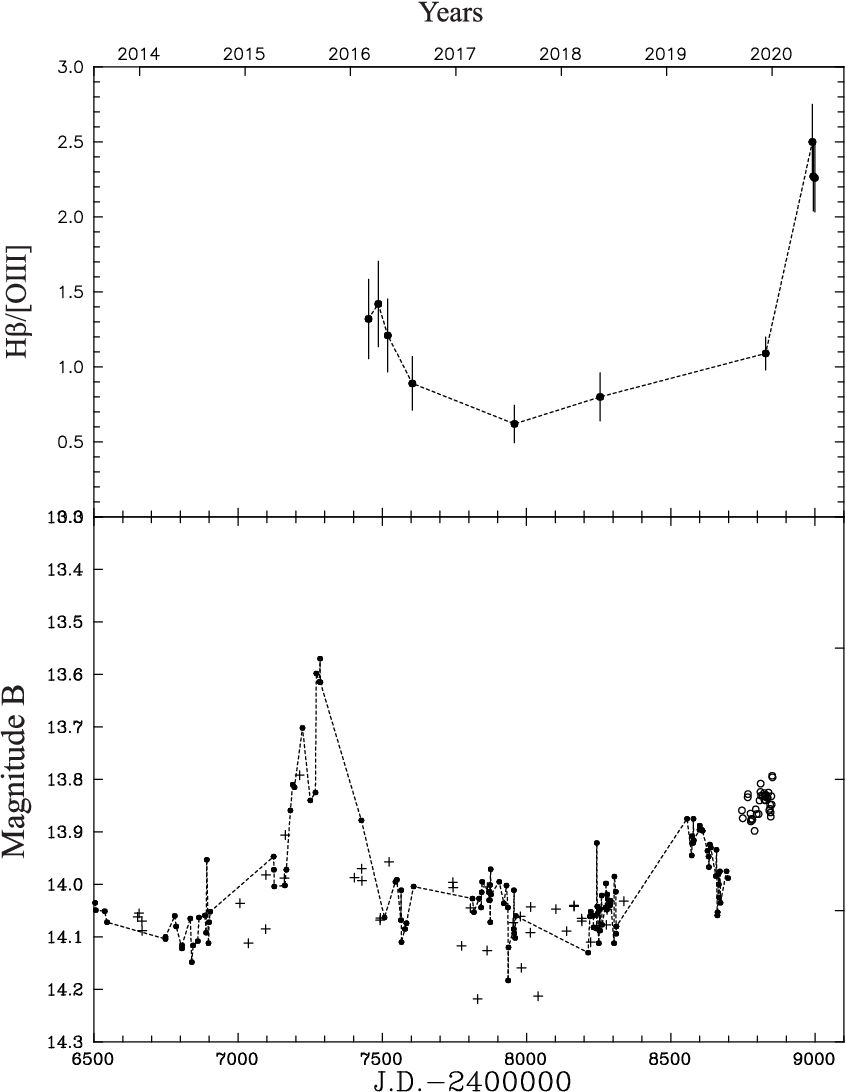}
\captionstyle{normal} \caption{Upper panel: variability of the H${\beta}$ intensity, normalized   to the [OIII]$\lambda$5007, in 2016-2020. Bottom panel: light curve of NGC~3516 in the $B$  with an aperture of 10 arc seconds according to published data: points - photometry at the Crimean Astronomical Station of SAI MSU, crosses - photometry carried out at SAO RAS, circles - photometry carried out at CMO SAI MSU (see text).}
\end{figure}

\begin{table}[p]
\setcaptionmargin{0mm} \onelinecaptionsfalse
\captionstyle{flushleft} \caption{Information about the obtained spectra (see text)}
\bigskip
\begin{tabular}{|c|c|c|c|c|c|c|c|c|}
\hline
  Date&  MJD& Aperture
 &Resol.&Disp.&S/N&Exp.&Spectraph \\
 
  &  & (")
 & (\AA)&(\AA/pix)&&(s)&\\
\hline
04.03.16&57452 &2.4x12 &10.6 &5.3&70
& 450 & 1\\
07.04.16&57486 &1.2x12 &5.3 &2.7&46
& 600 & 1\\
10.05.16&57519 &1.2x30 &5.3 &2.7&53
& 600 & 1\\
03.08.16&57604 &1.2x30 &5.3 &2.7&36
& 600 & 1\\
22.07.17&57957&1.1x40 &4.2 &2.1&30
& 900 & 2\\
23.07.17&57604 &2.2x40 &8.3 &4.2&28
& 600 & 2\\
16.05.18&58255 &2.2x80 &8.3 &4.2&45
& 2400 & 2\\
11.12.19&58586&2.2x80 &8.0 &4.0&66
& 1000 & 3\\
22.05.20&58992&2.2x80 &8.0 &4.0&150
& 1200 & 3\\
28.05.20&58998&2.2x80 &8.0 &4.0&150
& 1500 & 3\\
30.05.20&59000&2.2x80 &8.0 &4.0&150
& 1500 & 3
\\[1mm]
\hline
\end{tabular}
\end{table}

\end{document}